\documentclass[a4paper,11pt]{article}
\usepackage[a4paper,margin=1in]{geometry}
\usepackage{graphicx}  % needed for figures
\usepackage{dcolumn}   % needed for some tables
\usepackage{tabularx}
\usepackage{bm}        % for math
\usepackage{amssymb}   % for math
\usepackage{subcaption}
\usepackage{amsmath}
\usepackage{authblk}
\usepackage{threeparttable}
\usepackage{booktabs}
\usepackage{multirow}

\usepackage[colorlinks=true]{hyperref} %设置跳转

\newcommand \nldbd{$0\nu\beta\beta$ }

\begin{document}

%\widetext

\title{Optimization of convolutional neural networks for background suppression in the PandaX-III experiment}
\author[2]{Shangning Xia}
\author[2]{Suizhi Huang}
\author[2]{Kexin Xu}
\author[4]{Tao Li}
\author[1,3]{Xun Chen\thanks{corresponding author: chenxun@sjtu.edu.cn}}
\author[1]{Ke Han}
\author[1,2]{Shaobo Wang\thanks{corresponding author: shaobo.wang@sjtu.edu.cn}}

\affil[1]{INPAC and School of Physics and Astronomy, Shanghai Jiao Tong University, MOE Key Lab for Particle Physics, Astrophysics and Cosmology, Shanghai Key Laboratory for Particle Physics and Cosmology, Shanghai 200240, China}
\affil[2]{SJTU-Paris Elite Institute of Technology, Shanghai Jiao Tong University, Shanghai, 200240, China}
\affil[3]{Shanghai Jiao Tong University Sichuan Research Institute, Chengdu 610213, China}
\affil[4]{Sino-French Institute of Nuclear Engineering and Technology, Sun Yat-sen University, Zhuhai 519082, China}

\date{\today}
\maketitle
\abstract{
 The tracks recorded by a gaseous detector provide a possibility for charged particle identification.
 For searching the neutrinoless double beta decay events of
  $^{136}$Xe in the PandaX-III experiment, we
  optimized the convolutional neural network based on the Monte Carlo
  simulation data to improve the signal-background discrimination
  power. EfficientNet is chosen as the baseline model and the
  optimization is performed by tuning the hyperparameters. In
  particular, the maximum discrimination power is achieved by optimizing the channel number of the top convolutional layer. In comparison with our previous work, the significance of discrimination has been
  improved by $\sim$70\%.}

\section{Introduction}
\label{sec:introduction}

Neutrinoless double beta decay ($0\nu\beta\beta$) is an extremely rare
type of nuclear decay process, during which two electrons are emitted
from a nucleus with an atomic number of $Z$, resulting in a nucleus of
$Z+2$, without the emission of neutrinos. The discovery of such a
process would imply the neutrinos are likely Majorana particles~\cite{Agostini:2022zub}, and the violation of
lepton number conservation.

Many experiments have been carried out to search for the \nldbd with
different isotopes, such as $^{76}$Ge~\cite{GERDA:2020xhi,
  Majorana:2022udl}, $^{130}$Te~\cite{CUORE:2022jto}, and
$^{136}$Xe~\cite{EXO-200:2019rkq, KamLAND-Zen:2022tow, NEXT:2021dqj},
through the spectrum analysis around the $Q$-value, which is the
energy released by the decay and shared by the two emitting electrons.
The current lower half-life limits of the three isotopes are
2.3$\times 10^{26}$~yr~\cite{KamLAND-Zen:2022tow},
1.8$\times 10^{26}$~yr~\cite{GERDA:2020xhi}, and
3.2$\times 10^{25}$~yr~\cite{CUORE:2022jto} (90\% confidence level, or
C.L.), established by KamLAND-Zen, GERDA, and CUORE experiments
respectively. Its predicted half-life is longer if the mass order of
neutrinos is a normal hierarchy.  Therefore, necessary new experiments
with larger volumes and new technologies are being actively planned
and constructed to further improve detection sensitivity. Reducing the
background level in the energy region of interest (ROI) is one of the
key requirements for the discovery of such a rare process. The gaseous
detectors are capable to record event tracks, providing an additional
way to discriminate signals from backgrounds through extracting track
features, thus being actively studied in recent
years~\cite{Gomez-Cadenas:2019ges}. In general, the gaseous
detector-based experiments utilize the prominent Bragg Blob (BB)
feature of event tracks. Each end of a \nldbd track has a BB, in which
rapid energy loss in unit volume happens because of increased
differential energy loss (Bragg peak) and larger scattering angles
before an electron stops.  However, only one BB is on the ends of the
track for the single electron background.  The PandaX-III experiment,
which utilizes a high-pressure gaseous xenon time projection chamber
(TPC) to search for \nldbd decay of $^{136}$Xe~\cite{Chen:2016qcd}, is
one of them. The study of this type of detector is pioneered by the
Gotthard experiment~\cite{Luscher:1998sd}. Algorithms to extract the
different topological features of signal and background tracks within
gaseous detectors have been developed with the operation and planning
of new experiments\cite{Ferrario:2015kta, Galan:2019ake, Li:2021viv}.

Machine learning techniques, such as deep convolutional
neural networks (CNN), have been introduced into experimental particle
physics for data analysis and particle
identification~\cite{Carleo:2019ptp, Aurisano:2016jvx,
  MicroBooNE:2020hho}. Recently, methods of neural networks are
applied in \nldbd search for background and signal
classification. Promising results are obtained based on both
simulation data and experimental data in the NEXT
experiment~\cite{NEXT:2020jmz}.
In the PandaX-III experiment, a special type of CNNs is used to discriminate the \nldbd signal from the gamma backgrounds within the same energy window based on simulation, 
leading to an average improvement of 61.8\% on the efficiency ratio of $\epsilon_s/\sqrt{\epsilon_b}$~(3.82 to 6.17) in comparison with its design baseline~\cite{Qiao:2018edn}.

In this work, we present the optimization of the neural network to
improve the discrimination power based on the design of the PandaX-III
detector. As a subsequent work of Ref.~\cite{Qiao:2018edn}, several
improvements are made in the treatment of Monte Carlo (MC) simulation
and the optimization of the neural networks.  We also aim to
understand the process of extracting and enhancing the physical
characteristics of tracks in CNN, which serves as the basis for
optimizing the model and improving identification ability.  The
article is organized as follows. A brief introduction to the
PandaX-III detector is given in Sec.~\ref{sec:detector}. The data
preparation, including the simulation of \nldbd and background events,
as well as the post-processing of data are discussed in
Sec.~\ref{sec:data}. Detailed information on the selection and
optimization of the neural networks and the results are presented in
Sec.~\ref{sec:optimization}. At last, a summary is given in
Sec.~\ref{sec:summary}.

\section{The PandaX-III experiment}
\label{sec:detector}

\begin{figure*}[htb]
    \centering
    \includegraphics[width = 0.7\linewidth]{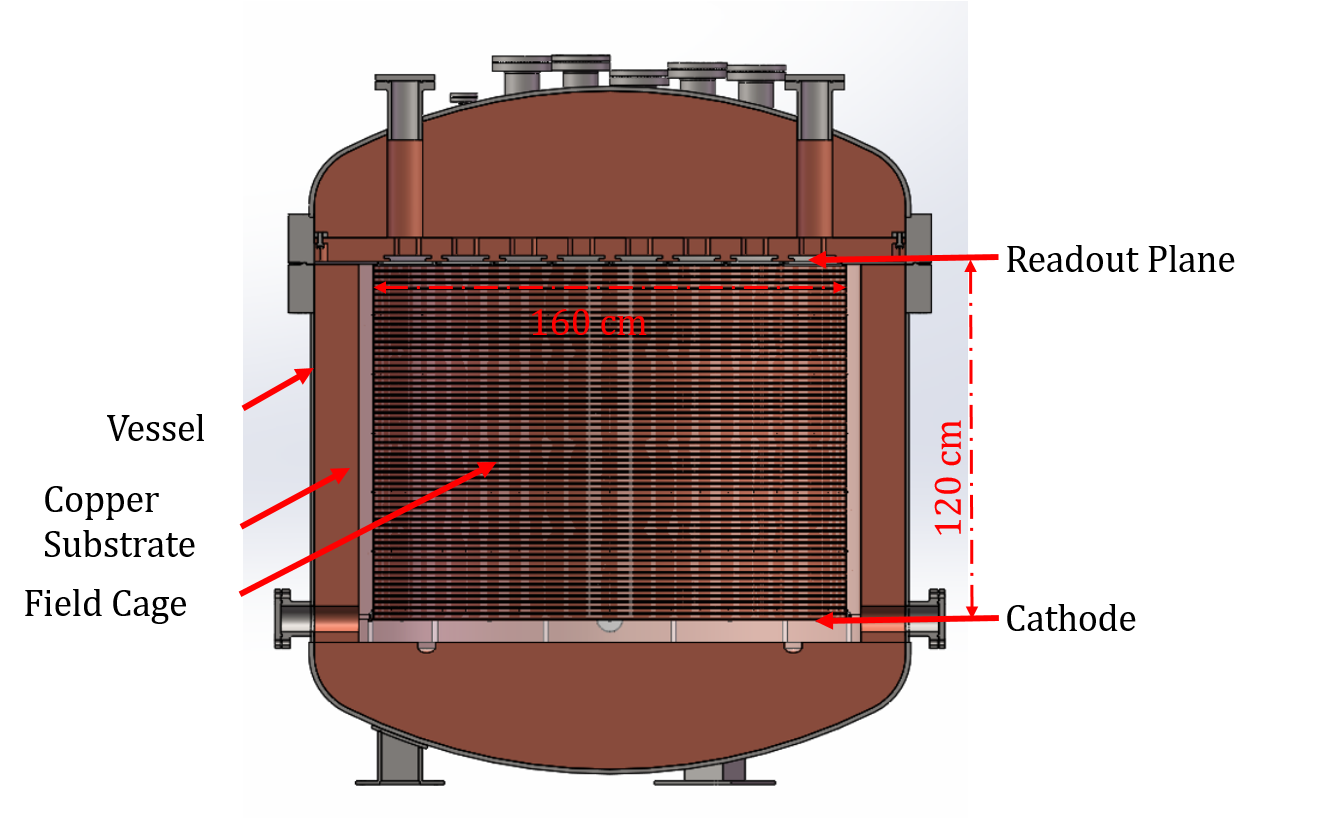}
    \caption{Illustration of the PandaX-III TPC from the cathode at the bottom to the charge readout plane with Micromegas on the upper side. All the main components are annotated.}
    \label{fig:pandax_iii_tpc}
\end{figure*}

The PandaX-III experiment under construction aims to search for \nldbd
with enriched $^{136}$Xe using high-pressure gas TPC in the China
Jinping underground laboratory (CJPL)~\cite{Kang:2010zza,Li:2014rca}. The updated
geometry of the PandaX-III detector is shown in
Fig.~\ref{fig:pandax_iii_tpc}, which is different from that in the CDR~\cite{Chen:2016qcd} and the previous work~\cite{Qiao:2018edn}. It has an active volume confined by the
field cage, cathode, and readout plane, with a drift length of 120~cm in the vertical direction~(Z)
and a diameter of 160~cm in the horizontal direction~(XY). It can contain $\sim$140~kg Xe gas with 90\%
enriched $^{136}$Xe, together with 1\% trimethylamine~(TMA) at the
operating pressure of 10~bar. The charge readout plane consists of 52
Micromegas modules~\cite{Andriamonje:2010zz}, each with a geometry of
$20\times20$~cm$^2$ and readout strips at 3~mm pitch. The cathode and
the electric field shaping cage are installed inside the copper
substrate with a width of 15 cm, which is used to shield the
radioactive background outside of the TPC. The materials have been
carefully selected to ensure low radioactive background inside the
TPC. More detailed information can be found in
Ref.~\cite{Wang:2020owr}.

The background for the search of \nldbd is from the radioactivities of detector construction materials ( $^{232}$Th, $^{238}$U, $^{40}$K, and $^{60}$Co).
Considering the energy distribution of the sources. The majority contribution is gamma from the
radioactive decay chains of $^{238}$U and $^{232}$Th, which may leave
energy depositions around the Q-value. The advantage of the gaseous
detector is to suppress these backgrounds with the recorded track features.

\section{Data preparation}
\label{sec:data}
Since there are no recorded data available, we use the simulated data to
study the methods of background suppression. The simulated data will
be converted to image format so that it could be accepted by the CNNs.

\subsection{Simulation and data production}

\begin{figure}[htb]
    \centering
    \includegraphics[width = 0.8\linewidth]{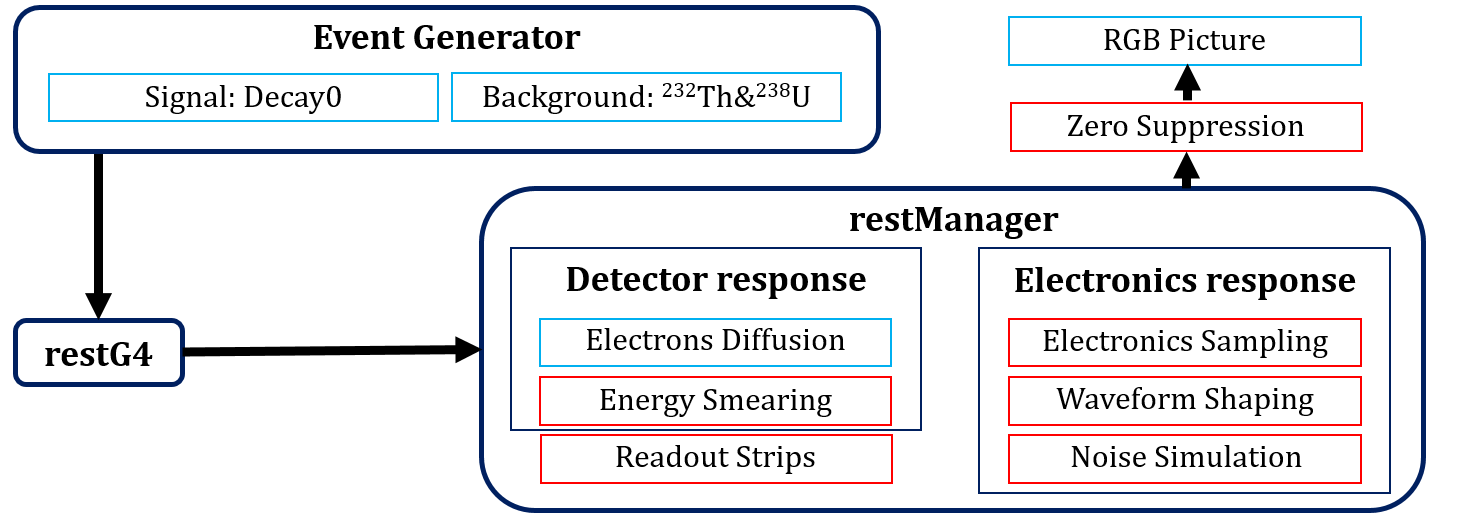}
    \caption{The process of data preparation. In comparison with data
      processing in Ref.~\cite{Qiao:2018edn}, new steps (in red
      frames) are added.}
    \label{fig:data_preparation}
\end{figure}

%The study of signal-background discrimination with CNN was carried out with Monte Carlo simulation events.
The Monte Carlo simulation is based on the detector geometry shown in Fig.~\ref{fig:pandax_iii_tpc}. The build of detector geometry in simulation is described in~\cite{Xie:2020xmd}.
All the components of the detector, including the vessel, the TPC and the dimension of the readout plane, are simulated.
The process of data preparation is shown in
Fig.~\ref{fig:data_preparation}. The physics simulation is performed
with the Geant4~\cite{Agostinelli:2002hh} based package restG4, which
is a component of the REST framework~\cite{Altenmuller:2021slh}. The
input signal \nldbd events are generated with the DECAY0
package~\cite{Ponkratenko:2000um}. The background events from the
decay chains of $^{232}$Th and $^{238}$U are sampled directly with
restG4. The output energy depositions inside the TPC are processed
with the REST framework, generating responses from the detector and
data readout by the electronics~\cite{Lin:2018mpd}. The readout data
is converted to images for further discrimination with the CNNs.

For each of the signal events, the DECAY0 package generates the momenta of
the two primary electrons, which share the same vertex. The vertices
are sampled uniformly in the gas region of the detector during the
simulation with restG4. The primary particles of the background events
are the daughter nucleus $^{208}$Tl ($^{214}$Bi) of the $^{232}$Th ($^{238}$U) decay chain, of which the 2615~keV~(2447~keV) gamma events contribute to the background within the ROI. They are sampled within the
container, which contributed most of these backgrounds. Since we have not yet determined the background model, we used the same uniform distribution of both signal and background events to study the effect on pure track features in this work.

\begin{figure}[htp]
  \centering
  \includegraphics[width=0.96\textwidth]{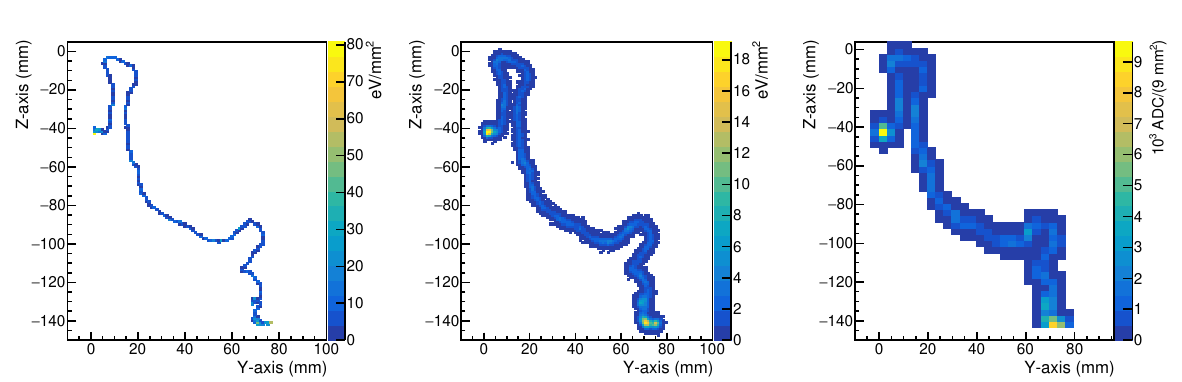}
  \caption{An example simulated \nldbd event at different processing stages by
   the REST framework, the projections on the YZ plane are shown. Left: The recorded energy
   depositions inside the PandaX-III TPC, simulated with restG4;
  Middle: The spatial distribution of ionized
   electrons reaching the readout plane; Right:
   The final spatial distribution of the readout signals.}
  \label{fig:xe136_event}
  \end{figure}

The restG4 package simulates the transport of the primary particles in
the materials as well as the daughter particles which are generated in different
physical processes. For each event, the energy loss of all the
particles inside the sensitive region of the detector, together with
the related positions, are recorded as energy
depositions. Fig.~\ref{fig:xe136_event} Left shows the recorded energy
depositions of an example \nldbd event in the TPC.

%\begin{figure}[hbtp]
% \centering
 %\begin{subfigure}[t]{.28\textwidth}
  % \centering
  % \includegraphics[width=1.0\linewidth]{figs/geant4.pdf}
  % \caption{energy deposition}
   %\label{fig:xe136_raw}
 %\end{subfigure}
 %\begin{subfigure}[t]{.28\textwidth}
  % \centering
   %\includegraphics[width=1.0\linewidth]{figs/diffusion.pdf}
   %\caption{electrons}
   %\label{fig:xe136_drift}
 %\end{subfigure}
 %\begin{subfigure}[t]{.28\textwidth}
  % \centering
   %\includegraphics[width = 1.0\linewidth]{figs/readout.pdf}
  % \caption{final signals}
   %\label{fig:final_signal}
 %\end{subfigure}
 %\caption{An example simulated \nldbd event at different processing stages by
   %the REST framework, the projections on the $y$-$z$ plane are shown: \subref{fig:xe136_raw}) The recorded energy
  % depositions inside the PandaX-III TPC, simulated with restG4;
  % \subref{fig:xe136_drift}) The spatial distribution of ionized
  % electrons reaching the readout plane; \subref{fig:final_signal})
  % The final spatial distribution of the readout signals.}
% \label{fig:xe136_event}
%\end{figure}

\begin{figure}[htp]
  \centering
  \includegraphics[width=0.6\textwidth]{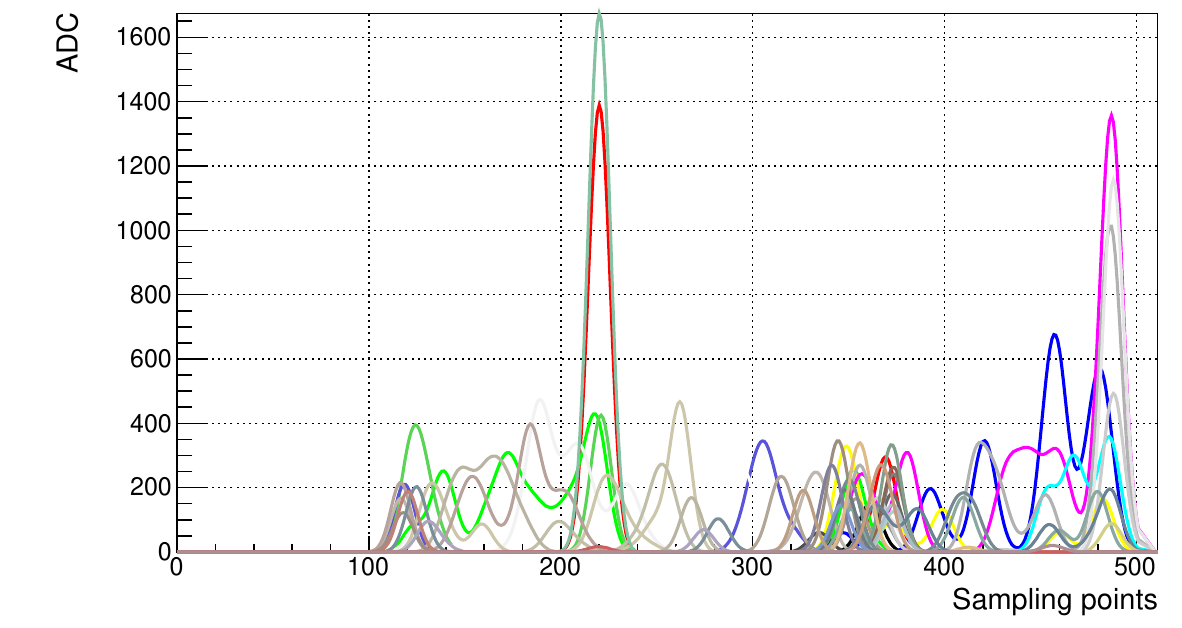}
  \caption{The generated waveforms of the example event, from
    the simulation of electronics response, noise and shaping by the
    REST framework, the waveform of the different strip is represented with different color.}
  \label{fig:waveform}
\end{figure}

The REST framework reads the energy depositions for further data
processing. For each of the energy deposition points, ionization
electrons are generated. The location distribution of the electrons (XYZ) is given by Geant4. The drift of the electrons to the readout
plane and their diffusion~\cite{NEXT:2019doo} are simulated. The parameters of drift
velocity and diffusion coefficient are calculated with the Garfield++
package~\cite{Garfield} integrated inside REST. With the designed
operation drift voltage of 1~kV/cm inside the TPC, the simulated
drift velocity is 1.87~mm/$\mu$s, and the transverse~(longitudinal)
diffusion coefficients are 1.0(1.5)$\times10^{-2}$~cm$^{1/2}$~\cite{Li:2021viv}.
Then the diffusion effect is simulated in both the transverse and longitudinal directions by performing a smear of Gaussian function according to the different locations of ionized electrons.
The horizontal position of each smeared electron is the projection on the XY readout plane, and the time arriving at the readout plane is calculated by applying the vertical location (Z) of each smeared electron and the drift velocity. Fig.~\ref{fig:xe136_event} Middle shows the spatial distribution of the
recorded electrons in the same example event on the readout plane by adding the diffusion effect,
where the $z$ distribution is concluded from the drift time.

The REST framework simulates the electron collection by the Micromegas and the electronics response. For the simulation data, the
REST framework groups the electrons to XY readout strips according to
their position, and creates an electronic response following the same
configuration used in the PandaX-III prototype
detector~\cite{Lin:2018mpd}, including a 5~MHz sampling rate, 5~$\mu$s
waveform shaping time, 120~fC dynamic range equivalent to 4096~ADC from electronics,
and 0.3~fC electronic noise level obtained from our real data taking of 10~ADC. Therefore, the waveform for each readout strip is
built. Fig.~\ref{fig:waveform} shows the waveforms of the example event. The simulated waveforms are digitized with an interval between two sampling of 200~ns. The event window is chosen to be 102.4~$\mu$s. Thus each event
has 512 time bins. The effective event window is 412 time bins which can ensure the full collection of $0\nu \beta \beta$ events, the first 100 bins are used for the baseline and noise study as well as in  the real data analysis.
The generated data has the same format as the data taken
in the real experiment. The spatial distribution of the final signals after the zero suppression of baseline and noise.
in the example event is shown in Fig.~\ref{fig:xe136_event} Right.
For the energy smearing, we did not simulate respectively the contributions from the avalanche process of Micromegas, gain non-uniformity of the detector, and the electron lifetime effect.
The amplitudes of energy depositions are smeared by a Gaussian
function to achieve the final design index of the energy resolution of 3\% FWHM~(Full
Width Half Maximum) at the Q-value.

\begin{figure}[hbt]
  \centering
   \includegraphics[width=0.4\textwidth]{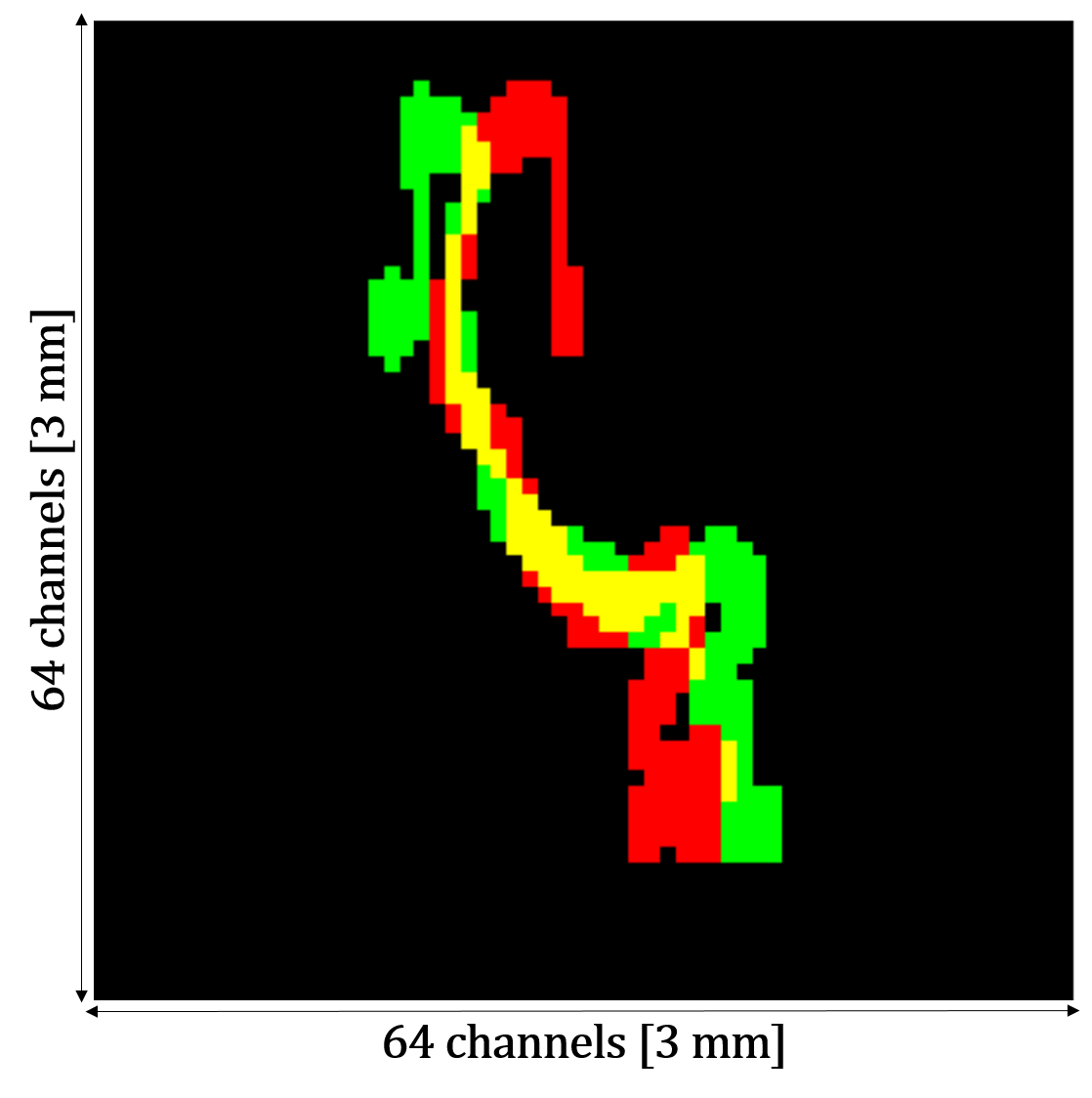}
   \caption{The result RGB image scaled from $64\times64$ pixels. The track projections in XZ and YZ planes are encoded in the red and green channels. All
     the pixels take the maximum values for better visualization.}
  \label{fig:res_image}
\end{figure}

\subsection{Data set}
\label{sec:data_set}
The generated events within the \nldbd search ROI of PandaX-III, from 2395~keV to 2520~keV, are selected to create the data set used for
the CNNs. The energy of tracks
on the XZ and YZ plane is normalized to the total energy of the event, and the tracks are then filled into red (XZ) and green (YZ) channels of the input
RGB image to the CNN model by aligning through charge centers of each plane.
As the strip size of the readout plane is 3~mm, in our study, the image format is $64\times64$ pixels with
a size of 3$\times$3~mm$^2$ each. Therefore, the image is produced  with the strip size in X or Y direction and is converted to the same size in the Z direction by calculating a drift distance using sampling points. The image size is 19.2~cm in each direction can ensure keeping the
whole track of \nldbd inside the image area. Fig.~\ref{fig:res_image} shows the resulting image of an example event.

The images then have been linearly scaled by a factor of 2 on each side to $128\times128$ by nearest neighbor interpolation before being fed into the network. The further discussion of this scaling is in Sec.\ref{sec3.2}.
Besides, to prevent over-fitting and improve the diversity of input data, data augmentations are applied, such as the random horizontal and vertical flipping of images.
These two methods involve horizontally or vertically flipping an input image with a certain probability during training phase.
Consequently, for different iterations, although the actual size of the dataset is unchanged, the same image fed into the network may be flipped or not.
By exposing the model to both the original and flipped versions of an event repeatedly, the model learns to recognize the same patterns from different viewpoints, leading to better generalization ability when presented with new, unseen events.

%\textcolor{red}{The images then have been linearly scaled by a factor of 2 on each side to $128\times128$ by nearest neighbor interpolation (one pixel is transformed into a square of 4 equal pixels), 
%before being fed into the network. The further discussion of this scaling is in Sec.\ref{sec3.2}.
%Besides, to prevent over-fitting and improve the diversity of input data,
%real-time data augmentations: random horizontal and vertical flipping have been applied.
%These two methods involve horizontally or vertically flipping an input image with a certain probability during training phase.
%Consequently, for different iterations, although the actual size of the dataset is unchanged, the same image fed into the network may be flipped or not.
%By exposing the model to both the original and flipped versions of an event repeatedly, the model learns to recognize the same patterns from different viewpoints, leading to better generalization ability when presented with new, unseen events.}

%The geometry of the result image is $64\times64$, corresponding to an area of $19.2\times19.2$~cm$^2$ on the readout plane. The conversion enables the whole track of a \nldbd event to be contained in the image.

%For better visuality, the strength of the detected signal on each point is neglected.

Our data set contains a total of 320,000 images from the \nldbd events
and the same number of images from background events, within which half
are from $^{232}$Th, and another
half from $^{238}$U. All the events are within the ROI. The images are
divided into three sets in this study: $70\%$ for training, $20\%$ for
validation, and the remaining $10\%$ for the evaluation of the
performance of the networks.

\section{Optimization of CNNs}
\label{sec:optimization}

Numerous CNN architectures have been developed in recent years.
InceptionV2~\cite{Ioffe:2015ovl}, DenseNet~\cite{huang:2018densely}, and EfficientNet~\cite{tan:efficientnet} have been proven to achieve remarkably higher accuracy and efficiency in the image classification than ResNet~\cite{He:2015wrn},
which was used for the discrimination of background and signal events in the \nldbd search with the PandaX-III detector~\cite{Qiao:2018edn}.

ResNet is a convolutional neural network architecture that introduced the concept of ``skip connections'' to enable the training of very deep neural networks.
It allows information to flow directly from one layer to another without passing through all the intermediate layers.
These skip connections create shortcuts that enable the network to learn residual mappings or the difference between the input and output feature maps of a layer.
InceptionV2 and DenseNet also made some improvements based on this idea.
InceptionV2 uses a series of convolutional filters with different sizes in parallel and splits an $n\times n$ convolution into $1\times n$ and $n\times 1$ convolutions,
allowing it to capture features of different scales effectively and to reduce the number of parameters and computations required.
DenseNet uses a ``dense connectivity'' pattern, where each layer is connected to every other layer in a feed-forward fashion.
This helps to improve the flow of gradients and information through the network, enabling it to learn more complex features.
It also introduces a technique called ``bottleneck layers'', which makes the training faster by compressing the input feature maps first and expanding them thereafter.
This helps to reduce the overall complexity of the network while still maintaining its representational power.
These ideas are then inherited by EfficientNet which we will explain in detail in paragraph~\ref{sec3.2}.

Thus, these architectures provide a possibility to improve the capability of background
suppression further in the PandaX-III experiment, as well as other experiments with similar technology.

%Since the emergence of ResNet in 2016, new CNN architectures with better performance in the domain of image classification
%were created every year.
%In order to further excavate the potential of background suppression
%through the particle track features and to determine which model is the most adapted one, we studied the performance of some selected network architectures to choose the baseline,
%upon which we will proceed to fine-tuning its parameters to obtain the optimal result. We first tested four different network models as shown in Table~\ref{comp}, of which the parameters and the training process are without proper tuning.
\subsection{Comparison of different CNNs}
\label{sec3.1}

To evaluate a network's performance for background suppression, the
most important metric is the significance $\Xi$, which is defined as
the highest ratio of $\epsilon_s/\sqrt{\epsilon}_b$ at different
selected thresholds of the output of the network, where $\epsilon_s$
and $\epsilon_b$ are the accepting efficiencies of signal and
background events, respectively. In the PandaX-III experiment, the
\nldbd detection sensitivity is linearly and positively correlated
with $\Xi$.

ResNet50, DenseNet169, InceptionV2, and EfficientNetB4
are selected in this study because they have a similar
level of trainable parameters and could be trained on our limited
computation resources. All the models are implemented with Tensorflow
~\cite{tensorflow2015-whitepaper} and trained with the
same generated data samples on two Nvidia Tesla V100 GPUs. The best
weights with the smallest validation losses are saved for the
classification of test data. The input shape of all the CNNs is
modified to accept images with a resolution of $128\times128$ and no other optimization is done.
The performance reference metrics are calculated with the test data and
given in Table~\ref{tab:comp}.
ResNet50 gives a comparable $\Xi$ with the result shown in \cite{Qiao:2018edn}, the tiny difference most likely originates from the detector geometry, a TPC with single readout plane is used in this work.
Additionally, for the baseline model selection, as the network models show a big difference for the signal-background discrimination, the fluctuation of each model is not further studied.
EfficientNetB4 gives the highest value
of the significance $\Xi$. Thus it is used as the baseline model for
further optimization.

\begin{table}[htb]
  \centering
  \begin{tabular}{cc}
    \hline
    {Network}      &  $\Xi$ \\
    \hline
    ResNet50			& 6.80	\\
    DenseNet169			& 7.92	\\
    InceptionV2			& 5.86	\\
    EfficientNetB4		&	9.80	\\\hline
  \end{tabular}

  \caption{Performance of the selected CNNs with the same test data set.}
  \label{tab:comp}
\end{table}

\subsection{Terminology}
\begin{itemize}
    \item Epoch: A complete pass through the entire training set during the training phase.
    \item Batch size: The number of samples sent into the neural network each iteration for the calculation of gradient descent optimization.
    \item Feature map: In CNN, a feature map is a 2D array of values that represents the output of a convolution operation between the input and the filter.
    \item Channel: Besides the width and the height of an input or output, the channel is the third dimension (e.g. a RGB image has 3 channels), which represents the number of feature maps produced by that layer.
    \item Receptive field: the receptive field refers to the area of the input image that a filter is influenced by. It represents the spatial extent of the input that contributes to the output of a single neuron. The receptive field tends to grow with the number of layers because each layer processes the output of the previous layer, incorporating information from a larger region of the input.
    \item Convolutional Layer: Like the mathematical definition of convolution, a convolutional layer is a layer of the neural network which has a certain number of filters. Each filter applies directly to the input and produces a feature map. Then, all the feature maps are stacked along the third dimension (channel) and constitute the final output of it.
    \item Max pooling: Max pooling is an operation that selects the maximum element from the region of the feature map covered by the filter, normally a $3\times 3$ or $2 \times 2$ square.
\end{itemize}

\subsection{Optimization of EfficientNetB4}
\label{sec3.2}
The structure of the baseline EfficientNetB4 model used in the
optimization is presented in Table~\ref{tab:effnetb4}. Its basic
component is called MBConv~(\textbf{M}obile inverted
\textbf{B}ottleneck \textbf{Conv}olutional
layer)~\cite{san:MobileNetV2, tan:MnasNet}, which employs the
depth-wise separable convolutional layer along with the
SE~(\textbf{S}queeze-and-\textbf{E}xcitation) module~\cite{8578843}.
The depth-wise separable convolutional layer separates the normal convolution operation into two consecutive ones.
First, it generates the output with the same channel number as its input, then, the output goes through a $1\times 1$ convolutional layer with the desired channel number.
By separating the spatial and channel-wise operations, depth-wise separable convolution can significantly reduce the number of parameters and computations required, while still maintaining or even improving the accuracy of the network.
SE module is a type of attention mechanism that compresses the input's spatial dimensions to a single channel
and learns the weights for each channel, resulting in the re-calibration of the output feature maps.
These structures could improve the performance of CNNs significantly compared to other networks we have mentioned in Sec.\ref{sec3.1}

\begin{table}
    \centering
    \begin{threeparttable}
        \begin{tabular}{ccccc}
            \toprule[0.4mm]
            {Stage} & {~~Operator~~} & {~~Input Shape~~} & {~~Output Shape~~} & {~~Repeat Count\tnote{*}}  \\
            \midrule[0.2mm]
            {1} & {Conv3$\times$3} & {128$\times$128$\times$3} & {64$\times$64$\times$48} & {1}
            \\
            {2} & {MBConv1, k3$\times$3\tnote{**}} & {64$\times$64$\times$48} & {64$\times$64$\times$24} & {2}
            \\
            {3} & {MBConv6, k3$\times$3} & {64$\times$64$\times$24} & {32$\times$32$\times$32} & {4}
            \\
            {4} & {MBConv6, k5$\times$5} & {32$\times$32$\times$32} & {16$\times$16$\times$56} & {4}
            \\
            {5} & {MBConv6, k3$\times$3} & {16$\times$16$\times$56} & {8$\times$8$\times$112} & {6}
            \\
            {6} & {MBConv6, k5$\times$5} & {8$\times$8$\times$112} & {8$\times$8$\times$160} & {6}
            \\
            {7} & {MBConv6, k5$\times$5} & {8$\times$8$\times$160} & {4$\times$4$\times$272} & {8}
            \\
            {8} & {MBConv6, k3$\times$3} & {4$\times$4$\times$272} & {4$\times$4$\times$448} & {2}
            \\
            {9} & {TopConv1$\times$1} & {4$\times$4$\times$448} & {4$\times$4$\times$n} & {1}
            \\
            {10} & {Pooling \& FC} & {4$\times$4$\times$n} & {1} & {1}
            \\
            \bottomrule[0.4mm]
        \end{tabular}
        \begin{tablenotes}
            \item[*] is the number of repetitions of the module in the network. The output of one module will be the input of another consecutively.
            \item[**] means this block's convolution layers have a kernel of size 3.
        \end{tablenotes}
    \end{threeparttable}
    \caption{Structure of the EfficientNetB4 baseline model. Each row
        describes a functional block with input shape and repeat
        count. The default $n$ of the last convolution is 1792.}
    \label{tab:effnetb4}
\end{table}

We altered the configuration and hyper-parameters of the chosen
baseline model to optimize its ability to extract physical features.
Since the original resolution of the images is $64\times64$, naturally we want to keep it unchanged when training the model. 
However, the formal input size of EfficientNetB4 is 380 which is a lot larger and even EfficientNetB0, its smaller version has an input size of 224. 
Choosing the best input size for the model is in fact reaching the compromise between time/memory cost and performance.
Furthermore, in order to not introduce any distortion of the events, the input sizes of 64, 128, and 256 are tested and the nearest neighbor interpolation is used for the resizing. The comparisons is shown in Table.~\ref{tab:compeff}. 
For EfficientNetB0, increasing the resolution from $64\times64$ to $128\times128$ can improve the significance of the model by 26\%, switching from B0 to B4 can further improve it by 17\%. 
At this configuration, it takes around 800 seconds for an epoch. However, further increasing the resolution to $256\times256$ costs 2.75 times more the total training time and double the memory usage with almost unchanged signal-background discrimination power.
Thus, 128 is set as the input size in this work.
The input and the output of the network have been modified so that it could accept the input image with the resolution $128\times128$ and generate a single value between 0 (background) and 1 (signal). As shown in Table~\ref{tab:effnetb4}, the default number of channels $n$ in TopConv is 1792.
Since average-pooling aims at smoothing the input by
taking the average of a region while max-pooling takes only the maximum value,
max-pooling will select those that it thinks are more important from a set of
features instead of combining them. This is exactly what we want the
model to do: to select and only select a couple of features that have physical meanings.
Thus, \textbf{max-pooling} is chosen as the pooling function before the final fully connected layer.
With learning rate set to be 0.001 and batch size set to be 256, our model fully utilizes the GPU's memory (occupies $\sim$32~GB), and takes about 9 hours for a training session.
To predict an unseen event, the computation time is only 1.46~ms in average.

\begin{table}
    \centering
    \begin{threeparttable}
        \begin{tabular}{cccc}
            \toprule
            {Network scale} & {~~Input resolution~~} & {~~Significance~~} & {~~Training time~~} \\
            \midrule
            {EfficientNetB0} &	{$64\times64$}   &	{6.66}	       & {~180s/epoch} \\
            {EfficientNetB0} &	{$128\times128$} &	{8.39 (+26\%\tnote{*})}  & {~350s/epoch (+94\%)} \\
            {\textbf{EfficientNetB4}} &	{\textbf{128$\times$128}} &	{9.80 (+17\%)}  & {~800s/epoch (+128\%)} \\
            {EfficientNetB4} &	{$256\times256$} &	{10.78 (+10\%)} & {~3000s/epoch (+275\%)} \\
            \bottomrule
        \end{tabular}
        \begin{tablenotes}
            \item[*] The percentages of improvement are compared with previous configuration.
        \end{tablenotes}
    \end{threeparttable}
    \caption{Cost and benefit of different network scale and image resolution.}
    \label{tab:compeff}
\end{table}

\begin{figure}[hbt]
  \centering
  \begin{minipage}[t]{0.48\textwidth}
      \centering
      \includegraphics[width=0.9\textwidth]{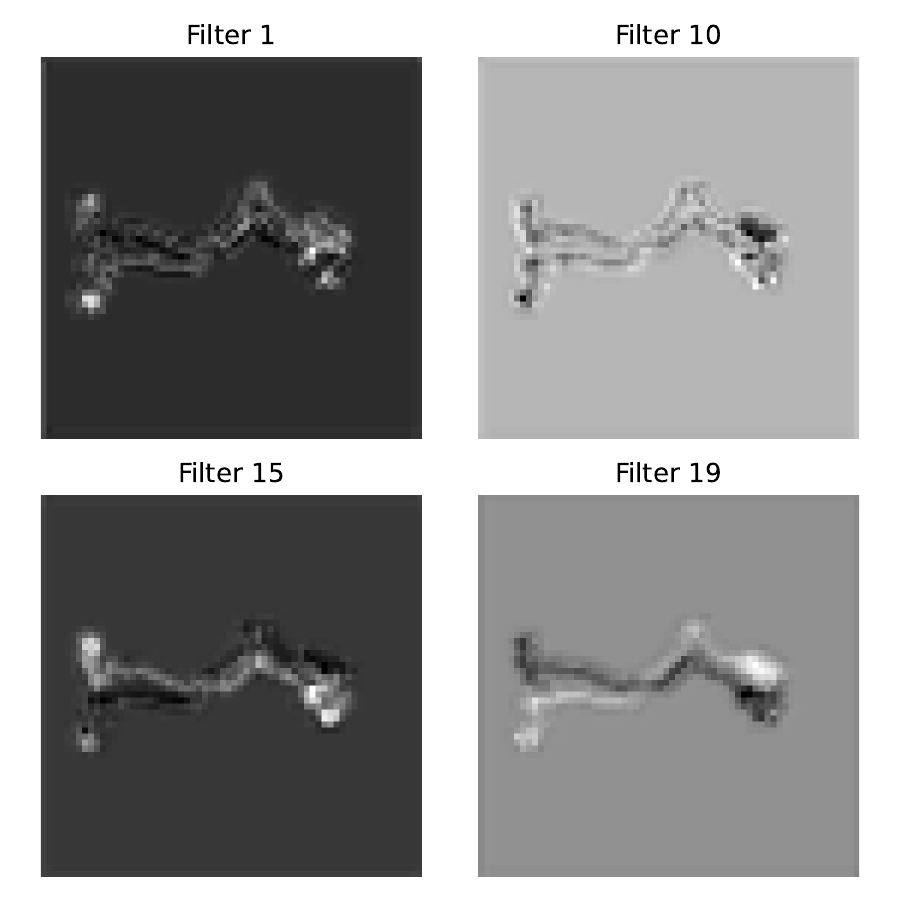}
      \caption{Example of the local feature extraction in Stage 2 (Table~\ref{tab:effnetb4}): Filter 1 detects energy depositions, Filter 10 detects contour of the tracks, Filter 15 and 19 detect the shape of projected tracks in YZ and XZ planes respectively.}
      \label{fig:layer1_example}
  \end{minipage}
  \hspace{.15in}
  \begin{minipage}[t]{0.48\textwidth}
      \centering
      \includegraphics[width=0.9\textwidth]{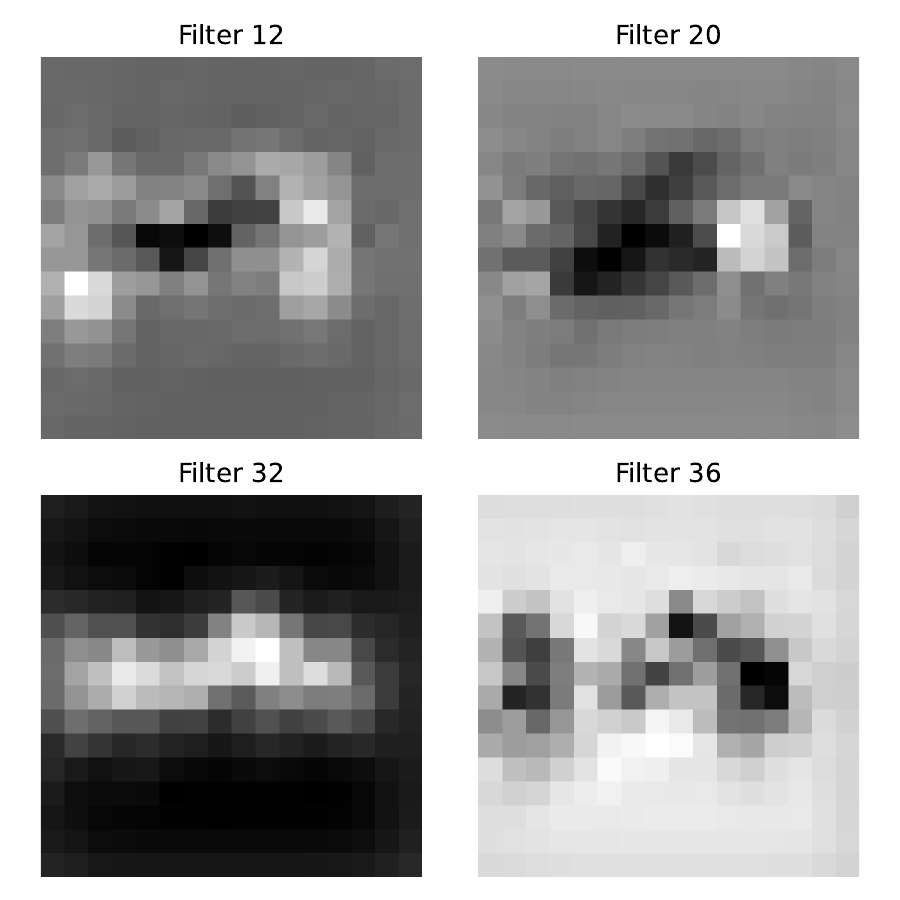}
      \caption{Example of the abstract feature extraction in Stage 4 (Table~\ref{tab:effnetb4}): Filter 12 and 20 detect energy deposition at the ends of the track, Filter 32 and 36 detect the geometric shape of the tracks.}
      \label{fig:layer3_example}
  \end{minipage}
\end{figure}

The convolutional neural networks are capable to learn the abstract
features encoded in the images~\cite{olah2017feature}.
The visualization of the feature extraction process in EfficientNetB4 is shown to understand how the model learns the local features and the high level abstract features of the events.
Lower layers is supposed to extract local features of the input images as their receptive filed is still small.
For example, Fig.~\ref{fig:layer1_example} shows four representative filters from the last layer in first MBConv block (Stage 2).
The complete feature maps of all 24 filters in this layer are presented in Appendix.~\ref{stage 2}.
The detailed and understandable local features are intuitively extracted, such as the energy deposition, the contour of the tracks, and the shape of projected tracks in YZ and XZ planes.
%We can see that they have covered what we intuitively think are important for the distinction of different events.
As the number of convolutional layers deepens, the local features are integrated and strengthened layer by layer.
In middle layers, for example in the forth layer (Stage 4) as shown in Fig.~\ref{fig:layer3_example}, the tracks begin to blur as resolution decreases.
But at the same time, the filters output enhanced physical features through integration of local features.
Some filters try to locate the energy depositions at the end(s) of the track (Brag peak) while others work on the geometric shape of the tracks.
The complete feature maps of all 56 filters in this layer are also presented in Appendix.~\ref{stage 4}.
These enhanced physical features are further integrated into high-dimensional physical features in subsequent layers.
It is worth noting that the function of a certain filter is consistent for different input, regardless a \nldbd event or a background event.

\begin{figure}[hbt]
    \centering
    \begin{minipage}[t]{0.48\textwidth}
        \centering
        \includegraphics[width=0.9\textwidth]{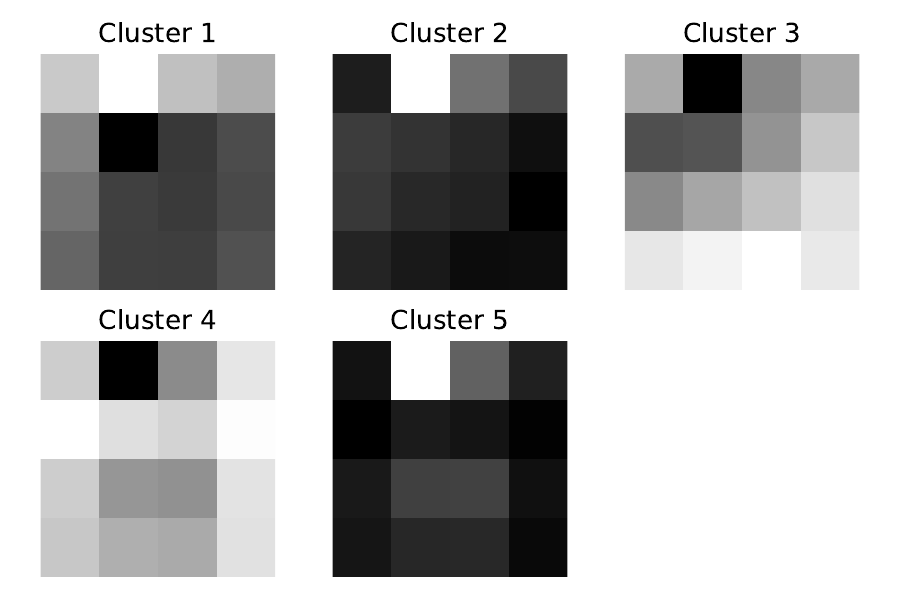}
        \caption{Clustering result of the feature maps from the last MBConv block. One example is selected for each cluster.}
        \label{fig:regroup}
    \end{minipage}
    \hspace{.15in}
    \begin{minipage}[t]{0.48\textwidth}
        \centering
        \includegraphics[width=0.9\textwidth]{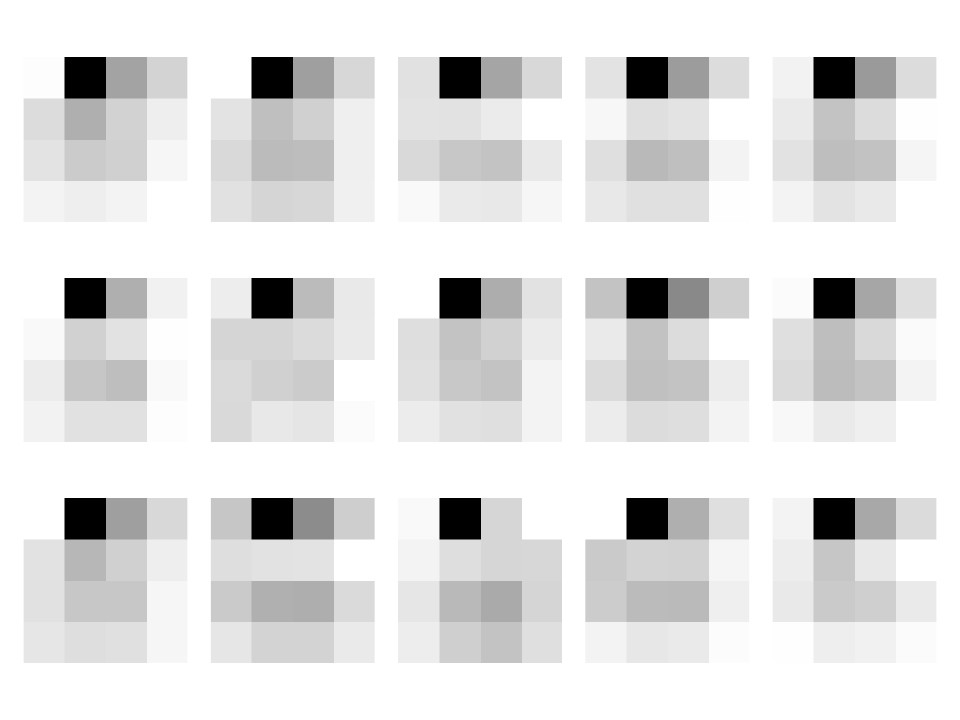}
        \caption{Feature maps of Cluster 4. There are 179  feature maps in Cluster 4, here randomly selected 15 feature maps for better visualization.}
        \label{fig:cluster}
    \end{minipage}
\end{figure}

The high-level features, which are used to direct determine the class of the input images, are obtained from the last layer of the last MBConv block (Stage 8).
The high-level features have become abstract and difficult to recognize directly the specific physical characteristics of the track.
However, these features originate from the local feature extraction and enhancement layer by layer in the model as shown previously in Fig.~\ref{fig:layer1_example} and Fig.~\ref{fig:layer3_example}.
Therefore,  the distribution of  high-level features distribution learned by the model is determined by the physical characteristics of events.
In this study, it is a reasonable conjecture that the number of key global high-level features and the physical characteristics for event classification have very close correlation.
The input images of tracks in this work are relatively less complex, thus the number of features needed for classification is limited.
The grouping of these high-level features is implemented with a widely-used clustering algorithm called DBSCAN~\cite{dbscan}, which clusters discrete hits based on proximity.
%similarities  to cluster them into different groups. Fig.~\ref{fig:regroup} shows the clustering results,
All the 448 filters of Stage 8 are automatically clustered into 5 groups (from Cluster 1 to 5) based on the similarity of feature maps as shown in Fig.~\ref{fig:regroup}.
The feature maps of the same group have extremely high similarity as presented in Fig.~\ref{fig:cluster}.
Therefore, a network could be then optimized by forcing the model to select a proper number of high-level features, or the
number of channels in the last convolutional layers.
In addition, $n$ could be interpreted as the number of possible features embedded in the image and the optimization could be achieved by choosing a proper $n$ corresponding to the need of the task.

\begin{figure}[hbt]
  \centering
  \includegraphics[width = 0.6\linewidth]{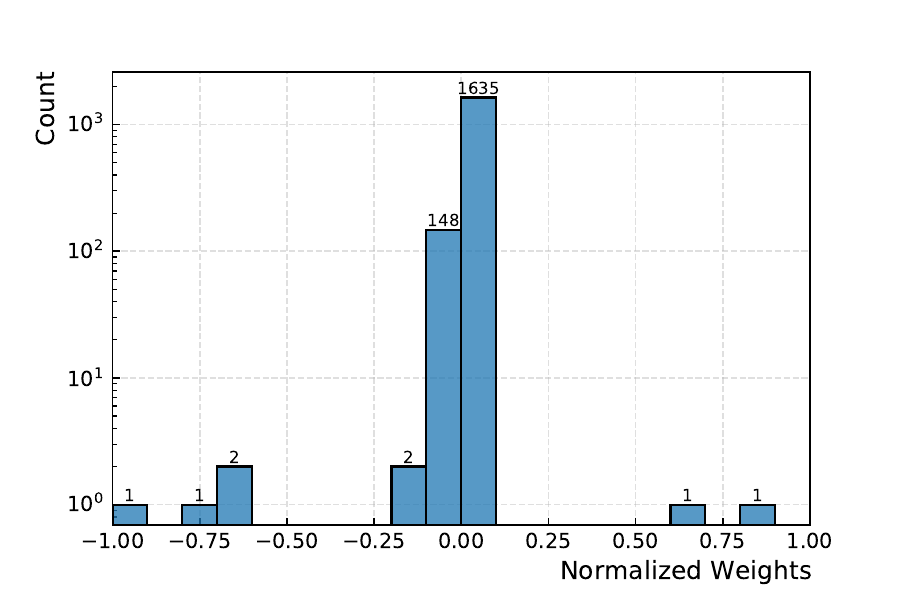}
    \caption{Distribution of the weight of the total 1792 channels in the fully connected layer.}
  \label{fig:weight_dis}
\end{figure}

At last, we analyze the weight distribution of FC (the fully connected layer in Stage 10).
The distribution is shown in Fig.~\ref{fig:weight_dis}, only 6 of the weights have an absolute value larger than 0.5,
including 2 positive ones and 4 negative ones. It suggests that only a
small subset of its input contributes significantly to the final
prediction value. To a certain extent,  it is also consistent with the visualization of the similarities clutering result of the high-level features. Thus, the principle optimization work is done by adjusting the channel
number $n$ in TopConv (Stage 9). The tested values are 2, 4, 6, 8, and 10.

To quantify the model's performance with a small prediction variance, we
calculate the final score using the following strategy:
\begin{itemize}
    \item only the epochs with validation accuracy are close enough ($delta<0.001$) to the maximum one are considered;
    \item for the qualified epochs, choose the weight from the first
      three epochs with the smallest validation loss values for prediction;
    \item the significances are calculated with the average prediction score from selected weights.
\end{itemize}

\begin{figure}[htb]
	\centering
	\includegraphics[width = 0.6\linewidth]{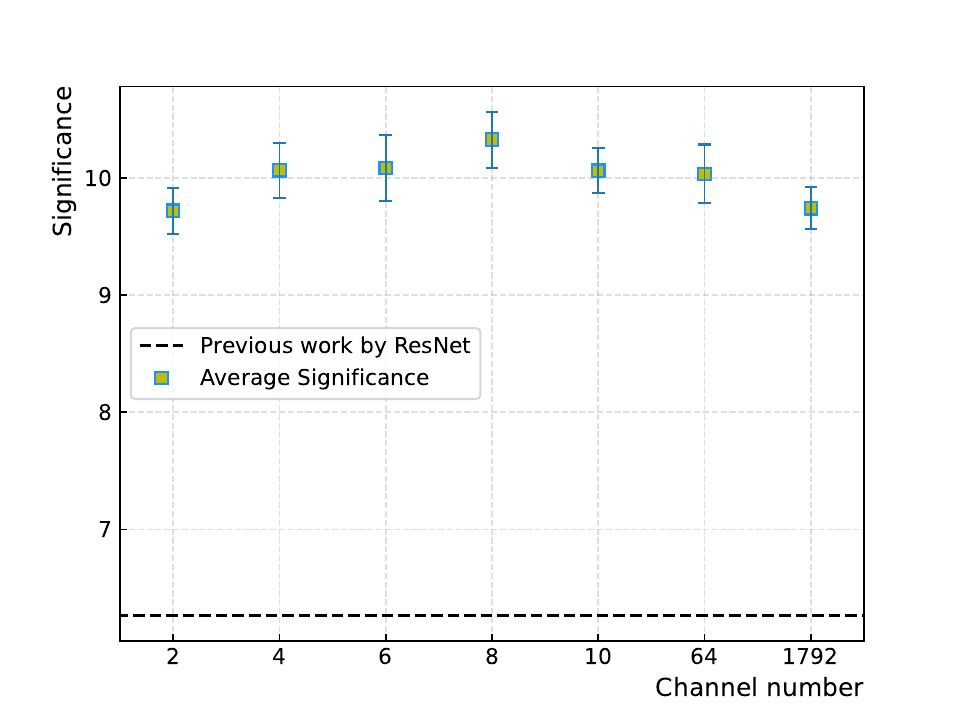}
	\caption{The significances for different models.}
	\label{fig:metrics_opt}
\end{figure}

The network is trained independently for six training sessions to
reduce the fluctuation for each value of $n$. The performance metrics
are shown in Fig.~\ref{fig:metrics_opt}. For the average significance
$\bar{\Xi}$, $n=1792$ and $n=2$ give the lowest values as
expected, as they either have an excess or insufficient number of channels. With a small but sufficient number of channels, such as 8, the significance is improved further by 10\% with an average of 10.32. Such significance is resulted from a highly suppressed background rate by 3 orders, with more than 35\% of signals retained.
% 30\% signal efficiency is retained with a background suppression efficiency of more than 3 orders of magnitude in the ROI.

Therefore, the result suggests that only a limited number of effective physical features
are embedded in the tracks of signal and background,
and that the matching between the channel number $n$ and the number of the effective features
can help to further improve the model's generalizability and performance.
Based on this assumption, we conclude that too many channels will not help improve the model's discrimination power. Meanwhile, too few channels will result in a lack of sufficient latent feature space
for the model to discriminate between the signal and the background events.

\section{Summary}
\label{sec:summary}

We studied the optimization of CNNs to improve the discrimination
power of signal and background events in the PandaX-III experiments,
based on Monte Carlo simulation data. The detector geometry is modeled
according to the latest design, and the responses of the detector and
electronics are simulated based on the actual parameters.  A more
realistic simulation dataset is generated.  EfficientNetB4 is selected
as a baseline model which shows a better performance among the models
studied.  We further explored the impact of network structure on the
discrimination ability for signal and background events. By checking the feature extraction and enhancement process of our neural network, the clustering of the high level features
and the weight distributions of the fully connected layer, we change the number of channels in the last
convolutional layer to match the possible physical features of the tracks.
The model with a small but sufficient number of channels in the last convolutional layer has a better performance,
which may be an indicator for the limited number of the physical features of tracks needed for signal-background discrimination.
A relatively high significance of $\sim$10 is obtained,
which is about 70\% higher than previous work~\cite{Qiao:2018edn}. Therefore, the direction of our future work is to establish a connection between the two and improve the interpret-ability of the neural network.

\section*{Acknowledgments}
This work is supported by the grant from the Ministry of Science and
Technology of China~(No.2016YFA0400302) and the grants from National
Natural Sciences Foundation of China~(No.11775142, No.11905127, and
No.12175139).  We thank the support from the Key Laboratory for
Particle Physics, Astrophysics and Cosmology, Ministry of Education.
This work is supported in part by the Chinese Academy of Sciences
Center for Excellence in Particle Physics~(CCEPP).

%\section*{References}
\bibliographystyle{unsrt}
\bibliography{mybibfile}
\appendix
\section{The features of Stage 2}
\label{stage 2}
\begin{figure}[hbt]
  \centering
  \includegraphics[width = 0.7\linewidth]{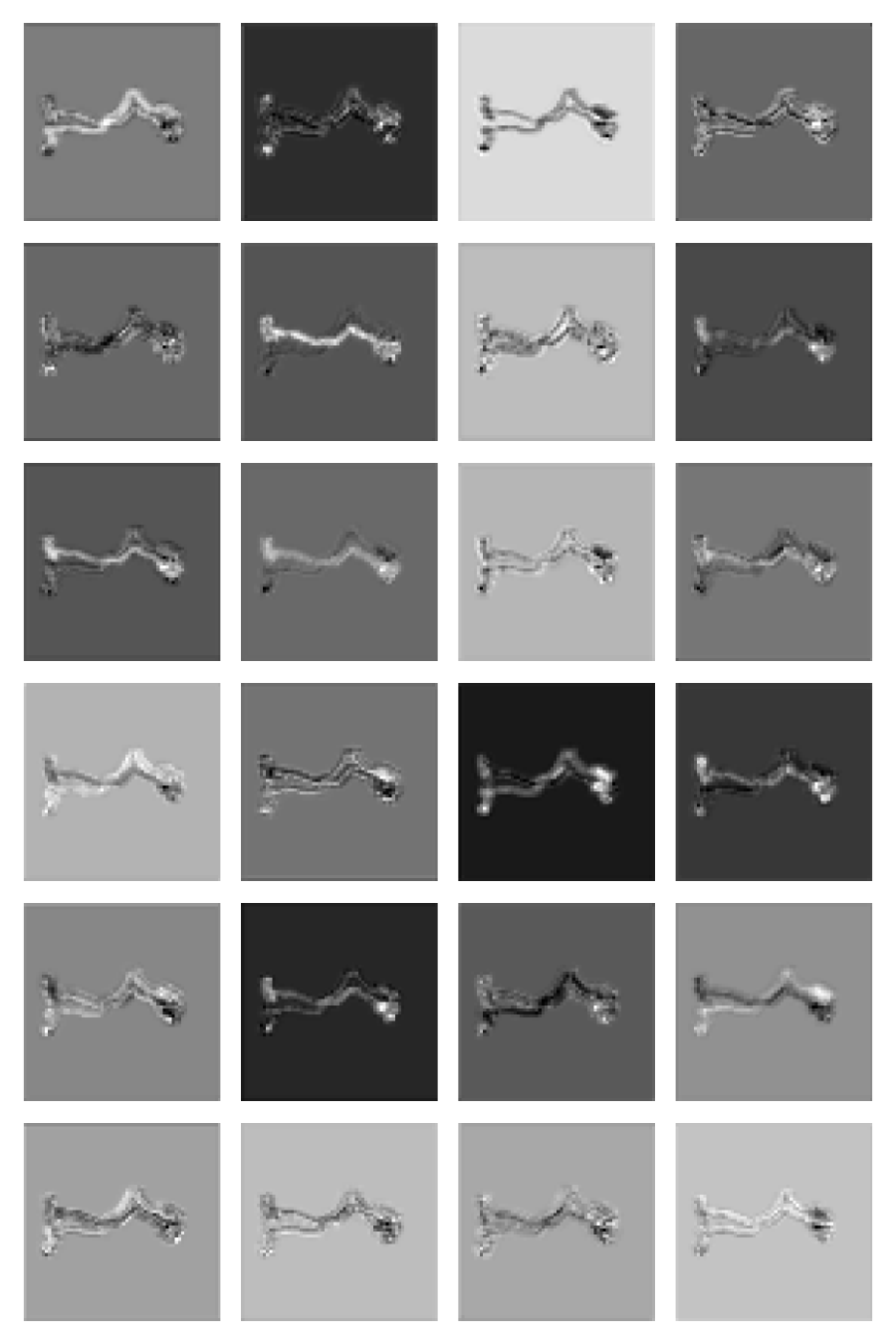}
    \caption{Example of the local features extraction in the first layer. All 24 filters are shown, and each filter extracts different local features of the tracks.}
  \label{fig:layer1_all}
\end{figure}

In the baseline model, the lower layers in the network learn local detailed features. For example, Fig.~\ref{fig:layer1_all} the 24 filters of the first layer.
The detailed physical features such as the energy deposition, the tracks in XZ and YZ planes, the geometric shape and edge of the track are extracted.

\section{The features of Stage 4}
\label{stage 4}

\begin{figure}[hbt]
  \centering
  \includegraphics[width = 0.9\linewidth]{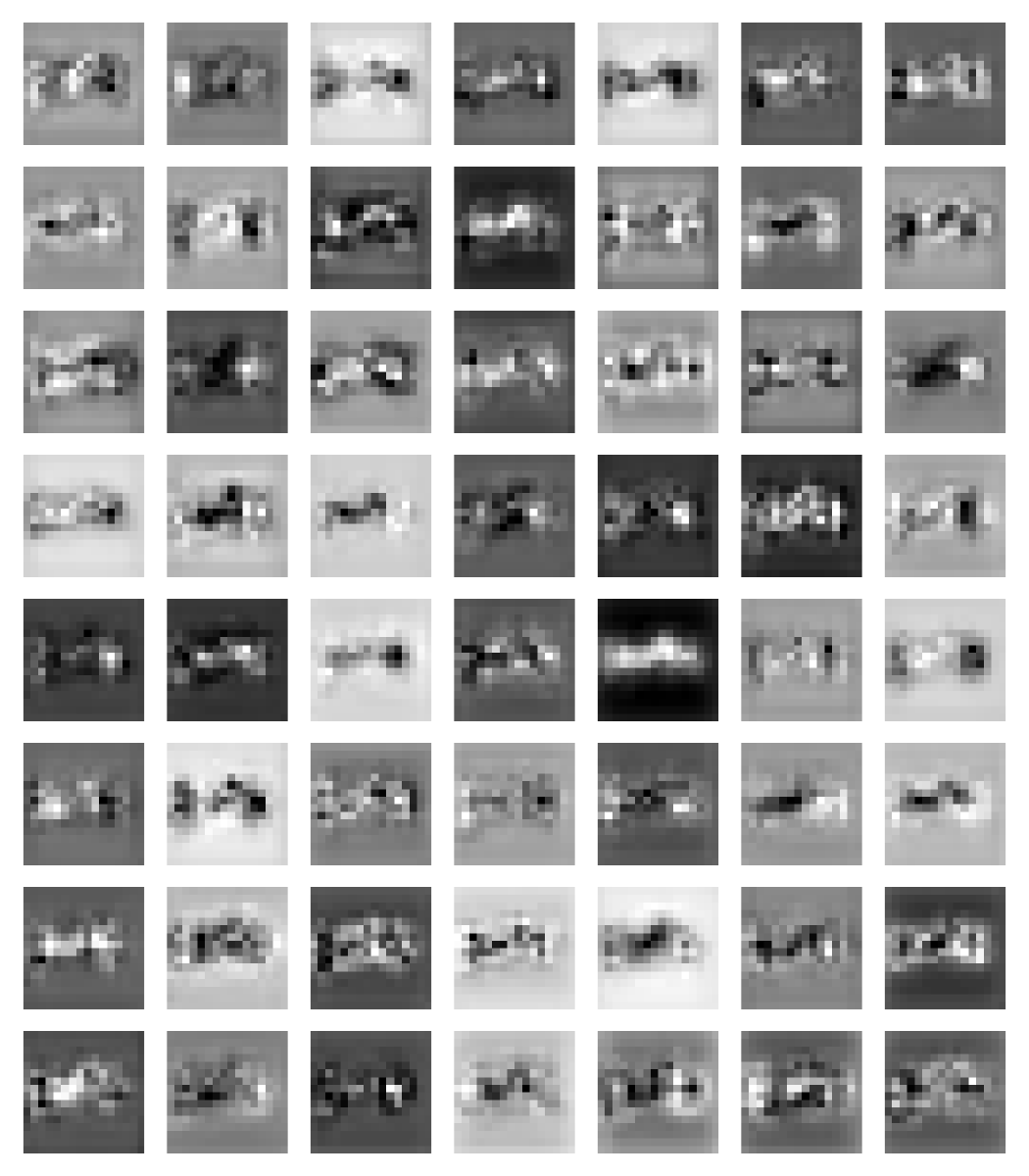}
    \caption{Example of the features enhancement in the third layer. All 56 filters are shown, and each filter tries to detect different enhanced local features of the tracks.}
  \label{fig:layer3_all}
\end{figure}

In the middle layers, the physical features are integrated and strengthened layer by layer. Fig.~\ref{fig:layer3_all} shows all the 56 filters of the third layer, the tracks begin to blur as resolution decreases.
The filters are activated locally on the tracks and the local features such as the energy deposition at the ends of the tracks, the geometric shape of the tracks are enhanced.

\end{document}